# Coherent phenomena in terahertz 2D plasmonic structures: strong coupling, plasmonic crystals, and induced transparency by coupling of localized modes

Gregory C. Dyer[a], Gregory R. Aizin[b], S. James Allen[c], Albert D. Grine[a], Don Bethke[a], John L. Reno[a], and Eric A. Shaner[a]

[a]Sandia National Laboratories, P.O. Box 5800, Albuquerque, New Mexico 87185; [b]Kingsborough College, The City University of New York, Brooklyn, New York 11235; [c]Institute for Terahertz Science and Technology, UC Santa Barbara, Santa Barbara, California 93106

**ABSTRACT**

The device applications of plasmonic systems such as graphene and two dimensional electron gases (2DEGs) in III-V heterostructures include terahertz detectors, mixers, oscillators and modulators. These two dimensional (2D) plasmonic systems are not only well-suited for device integration, but also enable the broad tunability of underdamped plasma excitations via an applied electric field. We present demonstrations of the coherent coupling of multiple voltage tuned GaAs/AlGaAs 2D plasmonic resonators under terahertz irradiation. By utilizing a plasmonic homodyne mixing mechanism to downconvert the near field of plasma waves to a DC signal, we directly detect the spectrum of coupled plasmonic micro-resonator structures at cryogenic temperatures. The 2DEG in the studied devices can be interpreted as a plasmonic waveguide where multiple gate terminals control the 2DEG kinetic inductance. When the gate tuning of the 2DEG is spatially periodic, a one-dimensional finite plasmonic crystal forms. This results in a subwavelength structure, much like a metamaterial element, that nonetheless Bragg scatters plasma waves from a repeated crystal unit cell. A 50% *in situ* tuning of the plasmonic crystal band edges is observed. By introducing gate-controlled defects or simply terminating the lattice, localized states arise in the plasmonic crystal. Inherent asymmetries at the finite crystal boundaries produce an induced transparency-like phenomenon due to the coupling of defect modes and crystal surface states known as Tamm states. The demonstrated active control of coupled plasmonic resonators opens previously unexplored avenues for sensitive direct and heterodyne THz detection, planar metamaterials, and slow-light devices.

## 1. INTRODUCTION

Interest in artificial electromagnetic structures with resonant elements much smaller than the wavelength of radiation with which they interact has remained high since the initial demonstrations of negative permittivity and permeability microwave metamaterials.[1, 2] The physical dimensions of subwavelength metal split ring resonators defined the '*LC*' resonances in these seminal metamaterial designs, and the operating frequencies of metamaterial devices were rapidly increased from microwave (GHz) frequencies into the infrared band[3] using geometric scaling principles. While researchers have devoted significant energy during the last decade to terahertz (THz) metamaterials[4, 5] due to the relative lack of technology in this frequency band, the first THz metamaterials can be traced back to efforts two decades earlier before the terminology 'metamaterial' had gained traction. Researchers at *Bell Laboratories* fabricated THz absorbers[6, 7] and emitters[8] in the late 1970s and early 1980s from Si metal oxide semiconductor field effect transistors (MOSFETs) and GaAs/AlGaAs heterostructures by patterning a quantum-confined two dimensional electron gas (2DEG) into macroscopic arrays of microresonators. In these 2DEG-based structures, two-dimensional (2D) plasma excitations where the physical dimensions determine the resonant plasmon wavelengths may be viewed as '*LC*' resonators. However, the equivalent circuit inductance and capacitance are distributed, in contrast to the lumped equivalent circuit description appropriate for metal-dielectric metamaterials. Broad interest in 2DEG structures as tailorable plasmonic micro- or nano-structures that provide effective macroscopic electromagnetic properties waned over the ensuing two decades, but the renaissance of low-dimensional electronic systems heralded by graphene and topological insulators has stimulated renewed efforts in this area.[9-15]

In this paper, we frame THz 2D plasmonics in a manner similar to metamaterials using an equivalent circuit transmission line model to describe our experimental studies. An integrated near-field detection scheme is used to characterize the spectrum 2D plasmonic structures. By employing this approach, elements composed of multiple voltage-controlled components that could form the primitive cells of a 2D plasmonic metamaterial array are studied. The subwavelength (relative to free space) nature of 2D plasma excitations enables the fabrication of plasmonic crystals at least an order of magnitude smaller than the wavelength of incident radiation. In our work, we explore not only the delocalized states that arise from the translational symmetry of the crystal lattice, but also the discrete modes that develop where translational symmetry is broken. A coupled resonator model provides a basis for phenomenological interpretation, and in conjunction with the equivalent circuit model accurately describes 2D plasmonic strong coupling and a phenomenon similar to electromagnetically induced transparency.

## 2. TRANSMISSION LINE MODEL OF 2D PLASMA WAVES

### 2.1 2D Plasmonic RLC Distributed Circuit Formalism

There are various methods to describe plasmons in low-dimensional electronic systems, including the random phase approximation,[16] hydrodynamic modeling,[17] and full field electromagnetic solutions[18, 19] in addition to transmission line modeling.[20] While each approach has advantages and disadvantages, a basic requirement fulfilled by all of these methodologies is the satisfaction of the 2D plasmon dispersion relation,

$$\omega^2(1 - i/\omega\tau) = \frac{n_{2D}e^2 q}{2m^* \epsilon_{eff}} . \tag{1}$$

Here $n_{2D}$ is the 2DEG density, $e$ is the electron charge, $q$ is the plasmon wavevector, $m^*$ is the electron effective mass, $\epsilon_{eff}$ is the effective dielectric constant of the semiconductor structure above and below the 2DEG, and $\tau$ is the plasmon scattering time. For the schematic in Fig. 1 (a), $\epsilon_{eff} = (\epsilon_b + \epsilon_a coth[qd])/2$ where $\epsilon_a$ is the permittivity of the semiconductor between the 2DEG at $z = 0$ and metal gate at $z = d$ and $\epsilon_b$ is the permittivity of the semiconductor below the 2DEG. We assume in this paper that $\epsilon_a = \epsilon_b = \epsilon_{GaAs}$, the permittivity of GaAs, consistent with the studied 2DEG structures.

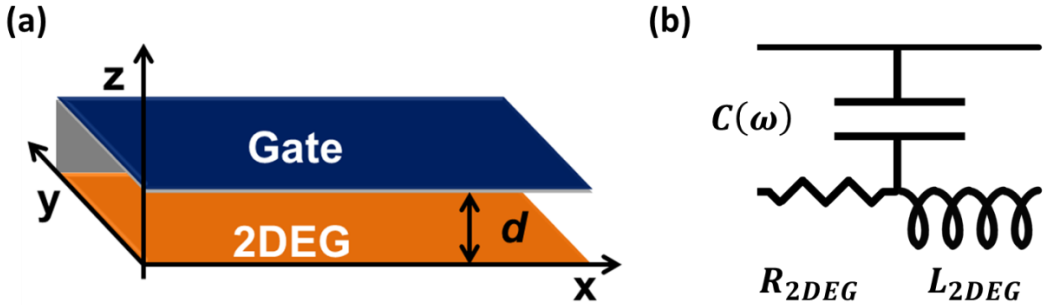

Figure 1. (a) A schematic of the 2DEG in a compound semiconductor heterostructure. The 2DEG is in the plane $z = 0$ with a metal gate located at $z = d$. The regions between the gate and 2DEG, $0 < z < d$, and below the 2DEG, $z < 0$, are comprised of the semiconductor materials surrounding the 2DEG. (b) The 2DEG in a transmission line equivalent circuit model is treated as a conductor confined to a plane with an inductance per unit length $L_{2DEG}$ and resistance per unit length $R_{2DEG}$ derived from the Drude conductance. The capacitance per unit length $C(\omega)$ depends on both frequency and the gate-2DEG separation $d$, though in the low frequency limit this capacitance reduces to a parallel plate capacitance.

Fig. 1 (a) can be represented as the equivalent distributed circuit illustrated in Fig. 1 (b).[20] The kinetic inductance per unit length $L_{2DEG} = m^*/n_{2D}e^2 W$ and resistance per unit length $R_{2DEG} = m^*/n_{2D}e^2\tau W$, where $W$ is the width of 2DEG, are derived from the Drude conductance such that $G_{Drude}^{-1} = i\omega L_{2DEG} + R_{2DEG}$. For typical 2D plasmon microresonator dimensions, the 2DEG kinetic inductance is several orders of magnitude larger than any inductance that arises due to the geometry of the structure. Thus, high mobility 2DEG plasmonic systems where $\omega L_{2DEG} > R_{2DEG}$ are kinetic inductance devices and support underdamped plasmons. The capacitance per unit length follows from a general definition of capacitance $C(\omega) = \delta n_{2D}/\delta \varphi = 2\epsilon_{eff} q$ where $\varphi$ is the electric potential in the plane of the 2DEG. Although the metal gate plays an important role in the plasmonic capacitance and its effects are accounted for in $\epsilon_{eff}$, even in the absence of a metal gate the capacitance $C(\omega)$ is properly defined and in fact is proportional to the plasmon wavevector $q$. Rigorous derivations of the 2DEG plasmonic transmission line model are in the literature,[20-22] but it is germane to note that this formalism is identical to the hydrodynamic description of 2D plasmons in the small signal limit where the hydrodynamic wave equations are linearized and is in the quasi-static limit.

The plasmon dispersion relation in Eq. 1 may be rewritten in terms of equivalent distributed circuit elements as,

$$iq = \sqrt{i\omega C(\omega)(i\omega L_{2DEG} + R_{2DEG})}. \tag{2}$$

Although this representation is no different symbolically from the standard transmission line model definition of the propagation constant, Eq. 2 is a transcendental equation except when $qd \gg 1$ or $qd \ll 1$. This complication arises because the 2D plasma wave is not generally a transverse electromagnetic (TEM) wave, but instead is a transverse magnetic (TM) wave. The wavevector dependence of the capacitance $C(\omega)$ is a direct byproduct of the presence of a longitudinal electrical field and its resultant displacement current. The plasmonic displacement current has a more significant impact of the equivalent transmission line impedance,

$$Z_{TL} = \frac{1}{\chi}\sqrt{\frac{i\omega L_{2DEG} + R_{2DEG}}{i\omega C(\omega)}}, \tag{3}$$

where

$$\chi = 1 - \frac{q(1-e^{-2q'd}\cos[2q''d])}{2q'(1-e^{-2q^*d})} + \frac{q e^{-2q'd}\sin[2q''d]}{2q''(1-e^{-2q^*d})} \tag{4}$$

and $q = q' + iq''$ with $q'$ and $q''$ denoting the real and imaginary parts of the plasmon wavevector. Though Eq. 4 is non-trivial in form, its physical significance is that it adjusts the characteristic impedance to account for energy carried by the displacement current. There are three field components contributing to the power flux in the system, two electric and one magnetic, but a transmission line approach treats the system as having only a voltage and current that produce transverse fields. This modification to the characteristic impedance preserves continuity of electric potential and conservation of energy by adjusting the relationship between the transmission line voltage $V_{TL} = \varphi$ and transmission line current $I_{TL} = \chi W J_{2DEG}$ such that $V_{TL} = Z_{TL} I_{TL}$ and $P_{TL} = Re[V_{TL}I_{TL}^*/2] = Re[Z_{TL}I_{TL}I_{TL}^*/2] = Re[V_{TL}V_{TL}^*/2Z_{TL}^*]$.[23] In this approach, the transmission line current $I_{TL}$ differs from the electronic current in the plane of the 2DEG, $W J_{2DEG}$, by the multiplicative factor $\chi$. The simplified treatment of the plasmonic fields circumvents complications that arise at boundaries and enables efficient calculation of the plasmonic modes that contribute to power flow in the system.

## 2.2 Calculation of Electromagnetic Fields

Although the transmission line approach determines the electric potential in the plane of the 2DEG and an effective transmission line current, it is nonetheless straightforward to calculate the two-dimensional field distributions using this one-dimensional model provided effects at boundaries and interfaces may be neglected. The general solution consists of forward and backward propagating voltage and current waves,

$$V_{TL}(x) = V_+ e^{-iqx} + V_- e^{+iqx} \tag{5}$$

and

$$I_{TL}(x) = (V_+ e^{-iqx} - V_- e^{+iqx})/Z_{TL}, \tag{6}$$

where $V_+$ and $V_-$ are complex amplitudes and $q$ is the complex plasmon wavevector found from Eqn. 1 or Eqn. 2. For the plasmonic system in Fig. 1, it follows that the electric potential $\Phi(x,z)$, electric field components $E_z(x,z)$ and $E_x(x,z)$, and magnetic field $H_y(x,z)$ are, respectively,

$$\Phi(x,z) = V_{TL}(x)\begin{cases} \sinh[q(d-z)]/\sinh[qd], z > 0 \\ e^{qz}, z < 0 \end{cases}, \tag{7}$$

$$E_z(x,z) = qV_{TL}(x)\begin{cases} \cosh[q(d-z)]/\sinh[qd], z > 0 \\ -e^{qz}, z < 0 \end{cases}, \tag{8}$$

$$E_x(x,z) = -iqZ_{TL}I_{TL}(x)\begin{cases} \sinh[q(d-z)]/\sinh[qd], z > 0 \\ e^{qz}, z < 0 \end{cases}, \tag{9}$$

and

$$H_y(x,z) = \frac{\epsilon_{GaAs}\omega}{c} Z_{TL} I_{TL}(x) \begin{cases} -\cosh[q(d-z)]/\sinh[qd], z > 0 \\ e^{qz}, z < 0 \end{cases}. \quad (10)$$

If boundary conditions are imposed as terminal impedances of an equivalent plasmonic transmission line circuit, then position dependent solutions to Eqns. 5 and 6 may be found and applied to Eqns. 7-10.

An example of this methodology is shown in Fig. 2. In Fig. 2 (a), the layout of the 2D plasmonic system is illustrated. Four gates 2.0 μm in width and separated by 2.0 μm are separated by 0.4 μm from the 2DEG by a dielectric with permittivity $\epsilon_{GaAs} = 12.9\,\epsilon_0$. Below the ungated regions, the 2DEG density is $n_0 = 4.0 \times 10^{11}$ cm$^{-2}$, while directly below the gates the 2DEG density is $n_1 = 0.24\,n_0$. Experimentally, the spatial non-uniformity of the gates is produced by application of a voltage to the gate terminals. The fields $\Phi(x,z)$, $E_z(x,z)$, $E_x(x,z)$, and $H_y(x,z)$ in this system with a 456 GHz excitation are plotted in Fig. 2 (b). Loss has been neglected for simplicity. Boundary conditions of $Z_{term} = 0$ were imposed at $x = 0$ and $x = 18$ μm, while the ungated and gate regions of the 2DEG were treated as discrete transmission line elements with step-like boundaries. For this particular excitation frequency and 2DEG density, the plotted solutions represent not only a resonance of the 18 μm long structure, but also resonances of the 2.0 μm subcavities formed by the gated and ungated 2DEG regions. Here the fundamental mode of a 2.0 μm plasmonic microresonator with wavevector $q_{fund} = 2\pi/4.0$ μm is evident in the ungated regions and the third harmonic of a 2.0 μm plasmonic microresonator with wavevector $q_{3rd} = 3q_{fund}$ is evident in the gated regions.

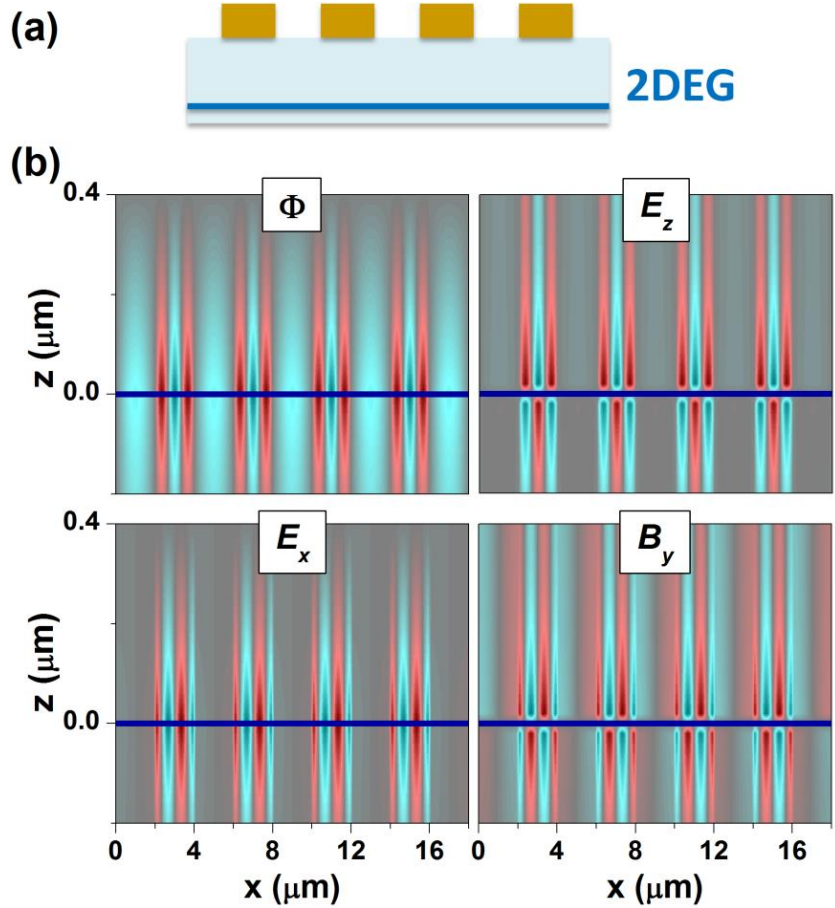

Figure 2. (a) A schematic of a 2DEG in a compound GaAs/AlGaAs semiconductor heterostructure 0.4 μm below a periodic gate with 4.0 μm period and 50% duty cycle. (b) The electric potential, transverse electric field, longitudinal electric field, and transverse magnetic field of a plasma excitation at 456 GHz where the 2DEG density below the gates is $n_1 = 0.24\,n_0$, where $n_0 = 4.0 \times 10^{11}$ cm$^{-2}$ is the 2DEG density in the ungated regions.

Several qualitative features in Fig. 2 merit additional discussion. The reduction in 2DEG density below the gates produces a plasmonic field enhancement relative the adjacent higher density regions.[24] Though all 2DEG regions are directly connected and equivalent circuit boundary conditions impose perfect coupling, the gated regions have relatively higher intensity fields and energy densities. Furthermore, the connection the gated regions through the ungated produces four coupled resonators in series, though the field distributions resulting from the matching of the plasmon wavelengths to the intrinsic gate dimensions only hint at the complex dynamics possible in this system. Finally, it can be seen that the longitudinal electric field is a non-trivial field component. Because the screening effects of the gate were assumed to be constant for both the gated and ungated regions of 2DEG, the significantly higher longitudinal field in the gated regions results from reducing the 2DEG density. The larger plasmon wavevector below the gated regions leads to a tighter confinement of the plasmon fields around the 2DEG, as seen in the lowering of the transverse electric field intensity moving upward from the 2DEG at $z = 0$ to the metal gate at $z = d$.

## 2.3 2DEG Plasmonic Crystals

The plasmonic structure diagrammed in Fig. 2 (a) may be interpreted as a four period plasmonic crystal where one set of adjacent gated and ungated regions of 2DEG form the crystal unit cell.[20-22] The crystal unit cell and the resulting multi-period crystal, which are both on the order of microns or tens of microns in size, are extremely subwavelength relative to the millimeter wavelengths of far infrared radiation in free space much like a metamaterial primitive cells. Yet the short plasmon wavelength results in plasma waves Bragg scattering from a repeated crystal unit cell. The resultant plasmonic band structure can be expressed in terms of a Kronig-Penny crystal dispersion relation[20, 25] using equivalent transmission line parameters,

$$cos[k_B(a_0 + a_1)] = cos[q_0 a_0]cos[q_1 a_1] - \frac{1}{2}\left(\frac{Z_0}{Z_1} + \frac{Z_1}{Z_0}\right)sin[q_0 a_0]sin[q_1 a_1] \qquad (11)$$

where $k_B$ is the Bloch wavevector and indices 0 and 1 correspond the ungated and gated regions of the unit cell, respectively. The characteristic impedances and propagation constants in Eq. 11 are found from Eqns. 2 and 3, while for the structure in Fig. 1 $a_0 = a_1 = 2$ µm.

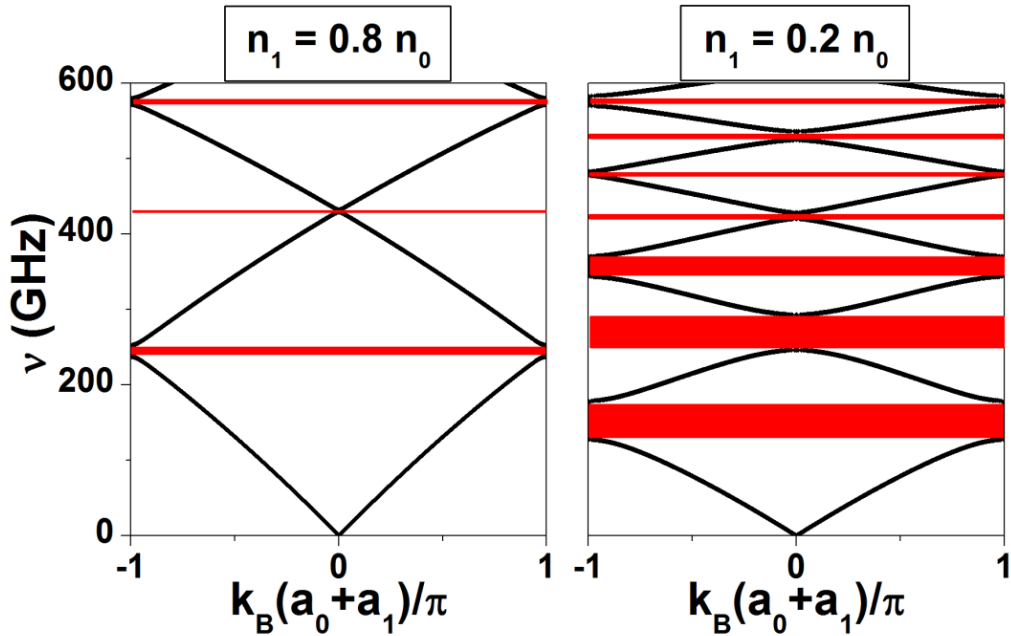

Figure 3. (a) Plasmonic crystal dispersion for the crystal unit cell of the structure in Fig. 2 for two different 2DEG densities in the gated regions of the 2DEG. The red areas are the plasmonic band gaps.

The infinite plasmonic crystal dispersion for the crystal unit cell of the structure in Fig. 2 is plotted in Fig. 3. Two configurations of the 2DEG density below the gate, $n_1 = 0.8\, n_0$ and $n_1 = 0.2\, n_0$, are considered. Because different 2DEG densities $n_1$ are accessible by application of a DC gate voltage, these two radically different band structures are readily accessible during operation of a single plasmonic crystal device. For example, the band edge below the first band stop region is found at 127.5 GHz with $n_1 = 0.2\, n_0$ and 237.9 GHz with $n_1 = 0.8\, n_0$, a greater than 50% *in-situ* tuning of the band edges using the band structure calculated with $n_1 = 0.2\, n_0$ as a reference. An interesting feature of this band structure is that the width of the band gaps is non-monotonic in frequency. Though at lower frequency or smaller values of $n_1$ the band gaps tend to be wider, there are certain frequencies where the band stops remain narrow or non-existent, independent of the broader trending of the band structure. These frequencies correspond to the scenario illustrated in Fig. 2 (b) where the ungated region of the plasmonic crystal unit cell resonates at 456 GHz. Evidently a resonance in the ungated, untuned region of the 2DEG, where $q_0(\omega) = j\pi/a_0$ and $j$ is an integer, suppresses formation of band stops and leads to gapless band structure when a higher and lower band intersect exactly at this resonant frequency. Also, since $q_0 a_0 = j\pi$ and $k_B(a_0 + a_1) = 0, \pi$ at these suppressed band gap points, it follows that $q_1 a_1 = l\pi$ ($l$ is an integer). Both elements of the crystal unit cell then must resonate independently despite their perfect coupling if the band gap vanishes between two bands. Interestingly, this effect is not unique to plasmonic crystals and occurs in one-dimensional photonic crystals. However, the continuous tunability of plasmonic crystal band structure makes these regions of gapless photonic transport more apparent.

## 3. COUPLED 2D PLASMONIC OSCILLATORS

### 3.1 Two Driven, Damped, Coupled Resonators

The recent experimental work of Dyer *et al.* on 2DEG plasmonic crystals has also examined the strong coupling of adjacent 2D plasmonic resonators[21] and the interaction of localized states in plasmonic crystals that result from breaking of translational symmetry.[22] While the transmission line model provides a precise description of the 2D plasmonic system, there is additional understanding that can be gained from describing this plasmonic system with a phenomenological coupled oscillator model. The equations of motion for two driven, damped, coupled oscillators in the time domain[26, 27] can be written as,

$$\begin{pmatrix} \frac{1}{\omega_a^2}\frac{\partial}{\partial t^2} + \frac{1}{Q_a \omega_a}\frac{\partial}{\partial t} + 1 & \kappa \\ \kappa & \frac{1}{\omega_b^2}\frac{\partial}{\partial t^2} + \frac{1}{Q_b \omega_b}\frac{\partial}{\partial t} + 1 \end{pmatrix} \begin{pmatrix} A(t) \\ B(t) \end{pmatrix} = \begin{pmatrix} F_a(t) \\ F_b(t) \end{pmatrix} \quad (12)$$

where $\omega_{a,b}$, $Q_{a,b}$, $F_{a,b}(t)$ are the resonant frequencies, resonator quality factors, and external driving amplitudes of oscillators $a$ and $b$ when in isolation, and $\kappa$ is the coupling coefficient between the oscillators. Assuming external excitations of the form $F_{a,b}(t) = f_{a,b}(\omega)e^{-i\omega t}$ and solutions of the form $A(t) = a(\omega)e^{-i\omega t}$ and $B(t) = b(\omega)e^{-i\omega t}$, the frequency domain amplitudes of the two oscillators are,

$$\begin{pmatrix} a(\omega) \\ b(\omega) \end{pmatrix} = \begin{pmatrix} \frac{D_b f_a(\omega) - \kappa f_b(\omega)}{D_a D_b - \kappa^2} \\ \frac{D_a f_b(\omega) - \kappa f_a(\omega)}{D_a D_b - \kappa^2} \end{pmatrix} \quad (13)$$

where

$$D_{a,b} = 1 - i\frac{\omega}{Q_{a,b}\omega_{a,b}} - \frac{\omega^2}{\omega_{a,b}^2}. \quad (14)$$

This general formulation describes a wide variety of coupled mechanical, electronic, and photonic resonator systems. The eigenfrequencies of the coupled oscillator system, assuming negligible loss, are given by,

$$\omega_\pm = \frac{1}{\sqrt{2}}\sqrt{\omega_a^2 + \omega_b^2 \pm \sqrt{(\omega_a^2 - \omega_b^2)^2 + 4\omega_a^2 \omega_b^2 \kappa^2}}, \quad (15)$$

and when the two (lossless) coupled resonators have identical resonant frequencies, $\omega_a = \omega_b = \omega_0$,

$$\omega_\pm = \omega_0 \sqrt{1 \pm \kappa}. \quad (16)$$

Although the coupling of two resonators with the same eigenfrequency $\omega_0$ lifts their degeneracy, the qualitative description of the underlying physics depends non-trivially upon the coupling strength, quality factors, and excitation amplitudes. Depending on the characteristics of the system, phenomena such as strong coupling,[28] Fano resonances,[29, 30] and electromagnetically induced transparency[27] may arise. Several cases will be considered generally in the next sections before examining experimental 2D plasmonic resonator systems in the context of this simplified coupled resonator model.

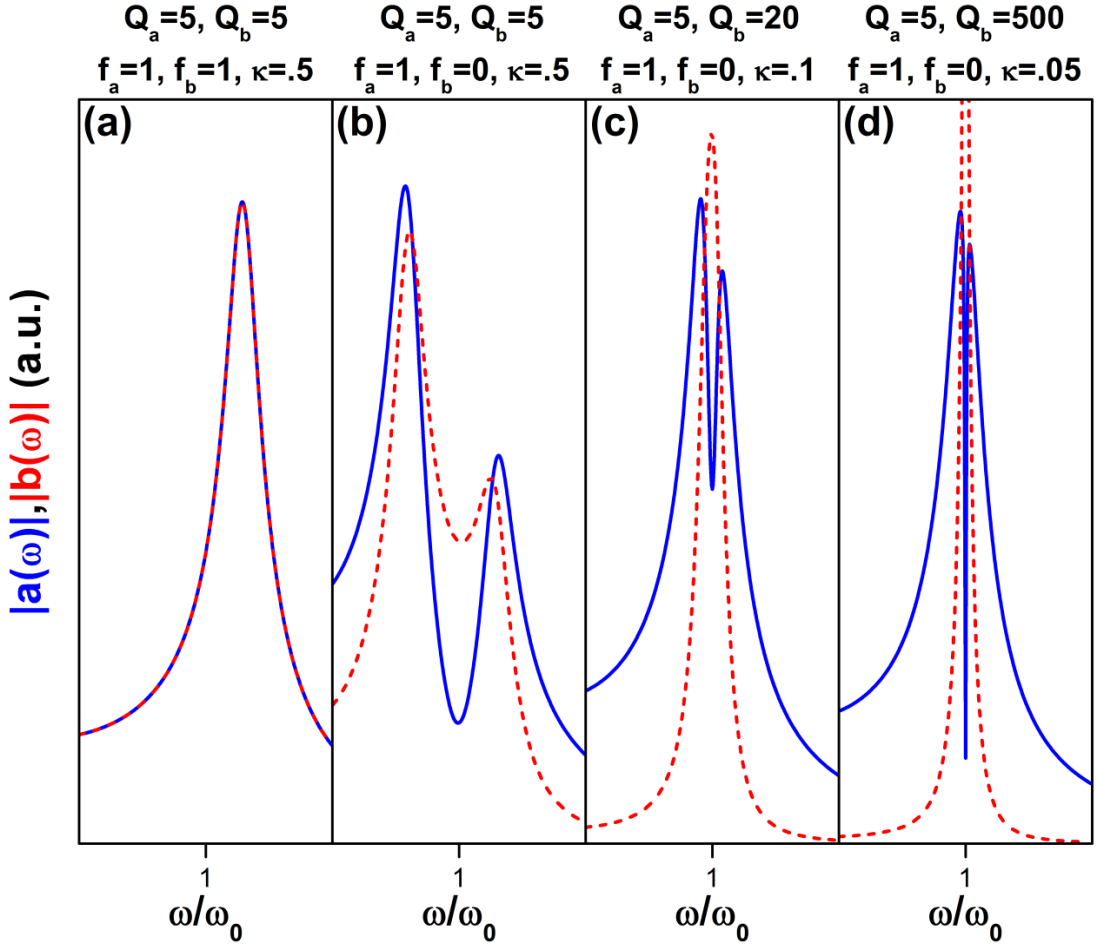

Figure 4. Absolute values of the amplitudes $|a(\omega)|$ (solid blue) and $|b(\omega)|$ (dashed red) of a pair of coupled oscillators $a$ and $b$ with identical resonant frequency $\omega_a = \omega_b = \omega_0$. The parameters used in Eqns. 13 and 14 are indicated above each pair of curves. The cases illustrated are (a) identically damped and driven oscillators, (b) identically damped oscillators, but with only oscillator $a$ driven, (c) non-identically damped oscillators with only oscillator $a$ driven, satisfying the minimal conditions for EIT, and (d) non-identically damped oscillators with only oscillator $a$ driven, satisfying the conditions for EIT in a nearly ideal system.

### 3.2 Identical Damped, Driven Oscillators

A trivial, but qualitatively instructive case concerns two coupled oscillators that are identical and also are driven identically. In this special case $\omega_a = \omega_b$, $Q_a = Q_b$, and $F_a(t) = F_b(t)$. It follows that $a(\omega) = b(\omega)$ in Eq. 13. When the frequency domain amplitudes of the oscillators are identical, then only the mode at the eigenfrequency in Eq. 16 with both oscillators in-phase, the symmetric mode, may be excited. In Fig. 4 (a), this case is illustrated and the absolute values of the amplitudes $|a(\omega)|$ and $|b(\omega)|$ overlap. The antisymmetric mode where oscillators $a$ and $b$ are 180 degrees out of phase has zero amplitude under these conditions. The presence of both modes from Eq. 16 in an analytical solution or experimental data thus is indicative of asymmetry in the quality factors, the external excitations, or both, between two coupled oscillators with the same resonant frequency, $\omega_a = \omega_b = \omega_0$. In Fig. 4 (b), the case where there is a strong asymmetry in the excitations of otherwise identical oscillators $a$ and $b$ is illustrated. A striking and important example of asymmetry in both quality factors and external coupling occurs in the classical analogue of electromagnetically induced transparency (EIT).[26, 27, 31, 32]

### 3.3 Electromagnetically Induced Transparency via Coupled Oscillators

The analogue of EIT, illustrated in Figs. 4 (c) and (d), in a classical coupled resonator system requires significant asymmetry between the resonators. While $\omega_a = \omega_b = \omega_0$ as in a generalized strongly coupled system, the conditions $|f_a(\omega)| \gg |f_b(\omega)|$ and $Q_b^{-2} \lesssim |\kappa|^2 \lesssim Q_a^{-2} \ll 1$ should also be satisfied for classical EIT. In this description, the larger driving amplitude of resonator $a$ is directly related to its lower quality factor. Resonator $a$ is described as a 'bright' or 'light' resonator because it couples more efficiently to an external excitation, and consequently also has greater radiative losses relative to resonator $b$ such that $Q_a > Q_b$. Oscillator $b$ is referred to as a 'dark' resonator, and has a higher quality factor because radiative damping is a less significant loss mechanism. The further constraints on the resonator quality factors and coupling strength ensure that there are two peaks in the absorption spectrum of the coupled oscillator system with a relatively small separation such that there is not a clear splitting into two modes as illustrated in Fig. 4 (c). In an ideal system with $Q_a \ll Q_b$, there is a very narrow minimum at $\omega_0$ in an absorption resonance centered at $\omega_0$ as shown in Fig. 4 (d).

EIT in its original form describes the opening of an optical transmission 'window' where there is strong absorption at the frequency an atomic transition line. EIT arises when one laser drives the aforementioned (radiative) transition and a second laser simultaneously drives a non-radiative transition in a three-level quantum mechanical system. This second laser generates a coherence whose strength is parameterized by its Rabi frequency and is proportional to the amplitude of the driving field. In the classical analogue, however, the requirement of two coherent photonic sources is circumvented by the intrinsic coupling between two oscillators. Thus, the coupling coefficient $\kappa$ in the classical picture plays the role of the Rabi frequency because it establishes coherence between the two resonators.

A brief discussion of the difference between Autler-Townes splitting and EIT in three-level atomic systems is helpful in further delineating the underlying physics.[33, 34] In Autler-Townes splitting, a single absorption line is split into two distinct modes due to a strong coherence and can be modeled as a pair of Lorentzians. However, EIT results from a weaker coherence and does not resolve the spectrum into a pair of widely split modes. Rather, the absorption spectrum has a Fano line shape where a discrete mode, a dark, narrow transition, interferes with a quasi-continuum, a broader, radiative transition. This interference between a direct radiative decay path and an indirect non-radiative path yields the characteristic relatively broad resonance with a narrow incision at its center. Taking both phenomena into account, generic strong coupling of classical oscillators is thus analogous to Autler-Townes splitting. This is seen in Fig. 4 (b) where both $|a(\omega)|$ and $|b(\omega)|$ have two peaks, although because $|\kappa| \sim Q_a, Q_b$ the splitting is on the same order as the resonator linewidths. In the EIT limit, on the other hand, $|a(\omega)|$ has two maxima at $\omega_\pm \approx \omega_0 \sqrt{1 \pm \kappa}$, while $|b(\omega)|$ has a single sharp peak located at $\omega_0$ as in Figs. 4 (c) and (d). It is this asymmetry in $|a(\omega)|$ and $|b(\omega)|$ that facilitates Fano interference, and oscillator magnitudes or power spectra with these qualitative features are a characteristic of EIT-like couling.

### 3.4 Coupled 2D Plasmonic Oscillators

For 2D plasmonic resonators with sufficiently large quality factors ($Q \gtrsim 5$), Eq. 2 may be used to define the 'characteristic frequency' of a plasmonic oscillator provided the resonance condition of the plasmon wavevector is determined by the geometry of a structure such as a gate.[22] The characteristic frequency is the resonant frequency of an isolated 2D plasmonic resonator of width $l$,

$$\omega_{cj} = q_c/\sqrt{L_j C_j} \qquad (16)$$

where $q_c = \pi/l$ and $L_j$ and $C_j$ are the equivalent transmission line inductance and capacitance, respectively, of 2DEG in the resonator. $L_j$ and $C_j$ are controlled by an applied gate voltage via tuning of the 2DEG density. Thus, even with a monochromatic excitation of a coupled resonator system, a pseudo-frequency domain representation of the 2DEG tuning may be used to provide a basis for comparison to a coupled oscillator model. For example, in Fig. 5 there are a series of resonators defined below *G2* and the four terminals of *G1* whose respective characteristic frequencies are $\omega_{ca} = \pi/2 \ \mu m \sqrt{L_a C_a}$ and $\omega_{cb} = \pi/2 \ \mu m \sqrt{L_b C_b}$. Here *G2* directly controls oscillator $a$, a defect that breaks the translational symmetry of the four period plasmonic crystal to its right, while oscillator $b$ is a localized plasmon mode known as a

Tamm state[20, 35] that arises when *G1* and the excitation frequency are tuned within plasmonic band gap. A detuning between these characteristic frequencies may be defined as $\delta\omega_c = \omega_{ca} - \omega_{cb}$.

The analogy between localized resonances in a system modeled by distributed circuit elements and discrete resonators has several limitations. The simple coupled oscillator model in the preceding sections very accurately describes two discrete resonators with a constant coupling coefficient. However, in the 2DEG system pictured in Fig. 5, there are two localized modes that can interact with one another as well as the delocalized states of the plasmonic crystal. Thus, there are generally more than two resonances to consider. Additionally, as the resonators are tuned, their quality factors and coupling change depending on their localization and radiative 'leakage' through boundaries. Finally, from an experimental standpoint, a time-average component of the complex plasmonic voltage adjacent to oscillator *a* (or *G2* as shown in Fig. 5) is measured by the mixing region below *G3*. Coupled oscillator effects may be observed in amplitudes or fields, but the loss (absorption) spectrum is what most rigorously defines EIT.

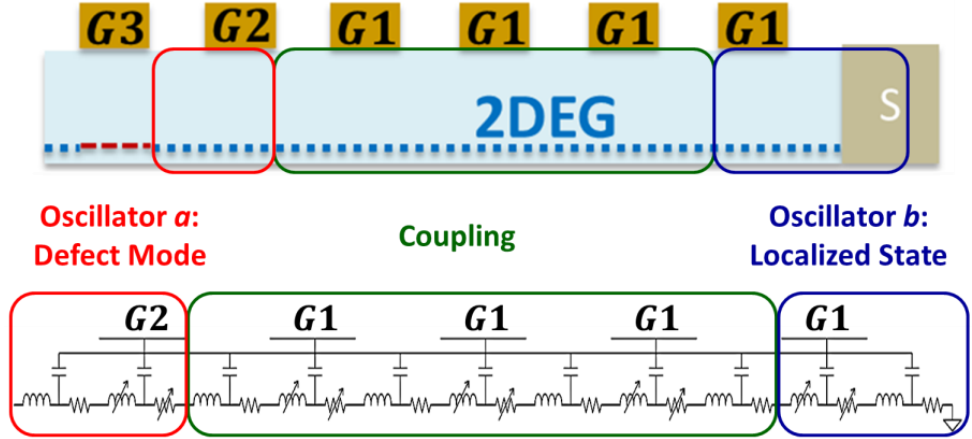

Figure 5. Schematics of a 2DEG-based plasmonic structure where the 2DEG is 0.4 μm below periodic gates with 4.0 μm period and 50% duty cycle. *G3* defines a mixing region below it, highlighted in red. *G2* and *G1* independently tune a multi-component 2D plasmonic structure between *G3* and S, an Ohmic contact. *G1* controls a four-period plasmonic crystal while *G2* tunes a single stripe of 2DEG. At excitation frequencies and gate tunings where the plasmonic crystal band structure produces a band gap, only localized states below *G2* and adjacent to S are permitted. In the equivalent circuit picture, these two effective oscillators are described by distributed *LC* resonators and are coupled through the bulk of the short plasmonic crystal lattice.

With these caveats in mind, we now consider recent experimental work (Dyer *et al.*, Nature Photon., 7, 925 (2013))[22] in the context of this coupled oscillator EIT model. The structure pictured in Fig. 5 was characterized at T = 8 K under 210.0 GHz excitation. A photovoltage resulting from plasmonic mixing was measured using a lockin technique, and was shown to be proportional to the THz 2D plasmonic voltage fluctuation at the right edge of the depleted region below G3. This experimental data is plotted in Fig. 6 (a) with the characteristic frequencies $\nu_{ca} = \omega_{ca}/2\pi$ and $\nu_{cb} = \omega_{cb}/2\pi$ and their detuning $\delta\nu_c = \nu_{ca} - \nu_{cb}$ parameterizing the tuning of gates *G1* and *G2*. The negative polarity signal is plotted such that relative minima are bright blue and relative maxima are dark red. Based on rigorous analysis of the detection mechanism,[22] it was found that the relative maxima of this negative polarity signal correspond to resonances. In this data, the band gap of the plasmonic crystal is in the region $260\ GHz < \nu_{cb} < 350\ GHz$, and therefore the crossing of the two resonances occurs in the plasmonic band stop region. Modes found in the band gap of a periodic structure are necessarily localized. Because translational symmetry is broken at opposite edges of the lattice, the two localized resonances must appear at the left edge by *G2* and at the right edge by the Ohmic contact labelled S in Fig. 5.

The localized plasmon mode at the left edge of the structure that is tuned by G2 is defined as oscillator *a*, while the mode at the right edge of structure is defined as oscillator *b*. The characteristic frequencies $\nu_{ca}$ and $\nu_{cb}$ of these two modes if completely uncoupled are found from the eigenmodes of a TL model[20] based on the equivalent circuit in Fig. 5 and are plotted as black lines in Fig. 6 (a). Qualitatively, it is evident that there is a repelled crossing between these two modes due to their evanescent coupling through the region between them, boxed in green in Fig. 5. Additionally, because oscillator *b* is only clearly observed when oscillator *a* comes into resonances with it, $\delta\nu_c \approx 0$, there is a strong asymmetry in the external excitation of the two resonances. Oscillator $a$ corresponds to a 'bright' resonator, while oscillator $b$ corresponds to a 'dark' resonator. The asymmetry in external driving as well as the relatively weak coherence between the two resonances satisfy several of the key criteria for observation of EIT.

Additionally, the calculation shown in Fig. 6 (b) of the real part of the voltage at the left edge of the structure using the plasmonic TL model reproduces all of these qualitative features. Here we have assumed a lumped excitation adjacent to oscillator $a$ where an antenna radiation resistance provides the excitation source impedance. Despite the seemingly restrictive assumption of a single, point-like excitation, the strong agreement between experiment and the TL model in Figs. 6 (a) and (b) further validates that the measured spectrum arises due to strong asymmetry in the resonator excitations from a distributed source. The non-vanishing signal in Fig. 6 (a) along $\nu_{cb} = 290\ GHz$ for $\delta\nu_c < -50\ GHz$ does indicate that oscillator $b$ has non-zero coupling to the 210.0 GHz excitation field. But the external driving is nonetheless much weaker than that of oscillator $a$.

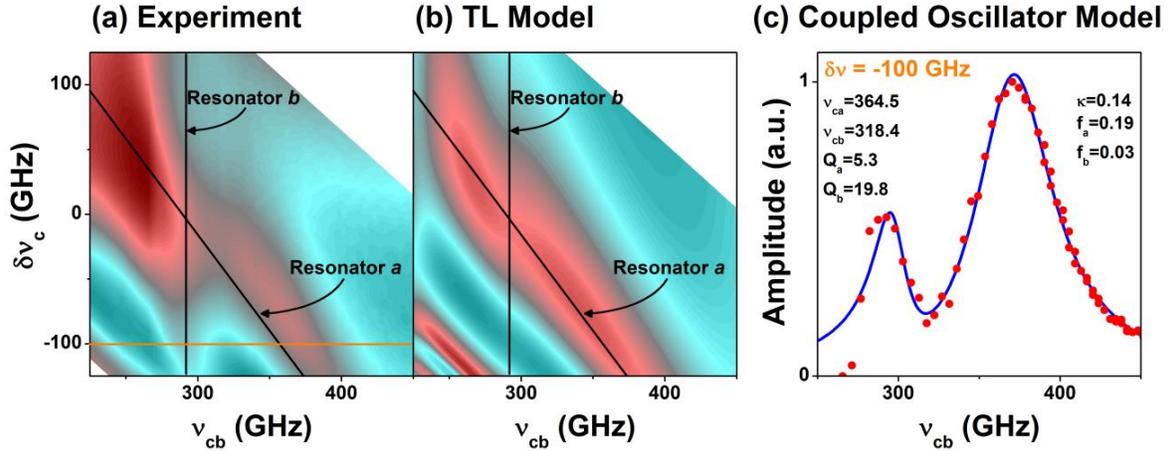

Figure 6. (a) A photovoltage measured across the 2D plasmonic structure illustrated in Fig. 5 is shown as a function of the characteristic frequency $\nu_{cb}$ of the 'dark' resonator $b$ and its detuning $\delta\nu_c$ from 'bright' resonator $a$. The excitation frequency is 210.0 GHz and the operating temperature was 8 K. The black lines indicate the position of the resonant characteristic frequencies of these two oscillators when their coupling is negligible. The horizontal orange line indicates where $\delta\nu_c = -100\ GHz$. (b) A TL model calculation of the photovoltage generated under 210.0 GHz excitation is shown for the 2D plasmonic structure diagrammed in Fig. 5. All parameters used in the model were measured experimentally or are taken from the literature. The color scales in both of the preceding figure parts are adjusted such that deep blue regions are relative minima and dark red regions are relative maxima in a negative polarity signal. (c) The self-normalized experimental photovoltage (red circles) at $\delta\nu_c = -100\ GHz$ under 210.0 GHz at T = 8 K. The data is fit to $Im[a(\omega)]$ (blue line), where $a(\omega)$ is defined by the simple coupled oscillator model in Section 3.1. The fit parameters are listed

Although many of the features of an EIT-like system are empirically satisfied by the coupled localizes states in this 2D plasmonic structure, additional analysis is needed to quantify the relationship between the coupling strength $\kappa$ and the resonator quality factors $Q_a$ and $Q_b$ in the context of EIT. The splitting is on the order of 10-20% of the center characteristic frequency at the crossing point, and from Eq. 16 the coupling coefficient is estimated to be approximately $\kappa \sim 0.1 - 0.2$. Although this magnitude of splitting is often characterized as strong coupling,[36] the coupling strength should be considered relative to the resonator quality factors for determining EIT. Here the splitting of the modes is of the same order as the linewidth of resonator $a$.

The broadening of 2D plasmonic resonances arises from several mechanisms. Drude damping and edge scattering are two intrinsic 2D plasmonic damping mechanisms that are independent of the external driving efficiency. Radiative damping rates, however, are often equal to or greater than Drude and edge scattering rates in cryogenic high mobility 2DEG systems.[37] Additionally, both radiative damping from an antenna coupled with 2D plasmonic resonator and radiative damping as a result of a 2DEG's induced dipole moment contribute in the studied system. For either of these broadening mechanisms, radiative damping can be described using a radiation resistance. Thus, both driving efficiency and radiative damping are directly connected. Strong coupling with an external excitation implies increased radiative losses, while poor coupling with an external excitation results in reduced radiative losses. A 'bright' 2D plasmonic resonance therefore is also heavily damped because of a good impedance match with its excitation source.

This directly impacts the resonator quality factor because radiative damping becomes a channel through which the resonator 'leaks' energy at one of its boundaries.

To better quantify the plasmon quality factors and their coupling strength, the experimental data has been fit to the coupled oscillator model discussed in Section 3.1. Data from Fig. 6 (a) with $\delta v_c = -100\ GHz$, highlighted by an orange line, was selected because the system can be approximated as having only two oscillators. For positive detunings $\delta v_c > 0$, it can be seen in both experimental and model plots in Fig. 6 that oscillator $a$ deviates from its uncoupled characteristic frequency. This is due to strong interaction with additional delocalized plasmonic crystal modes. Thus, data in this regime cannot be approximated by a pair of coupled oscillators and we instead focus on the negative detuning regime. The experimental data (red circles) and a fit of this data to $Im[a(\omega)]$ (blue line) are shown in Fig. 6 (c). Experimentally, only the real (or in-phase) component of a complex plasmonic voltage is measured. It should therefore be fit to either $Re[a(\omega)]$ or $Im[a(\omega)]$, and the imaginary component of the complex oscillator amplitude accurately describes the experimental data when a real driving amplitude is assumed. The data and its fit have a Fano lineshape rather than a more symmetric EIT spectrum because of the detuning of the resonances. Experimentally $\delta v_c = -100\ GHz$ while the fit finds that $v_{ca} = 364.5\ GHz$ and $v_{cb} = 318.4\ GHz$. However, the extracted quality factors, driving amplitudes, and coupling constant provide more vital information for determining if this system satisfies the requirements for EIT. Based on the extracted parameters shown in Fig. 6 (c), $|f_a(\omega)| \gg |f_b(\omega)|$ and $Q_b^{-2} \lesssim |\kappa|^2 \lesssim Q_a^{-2} \ll 1$ are fully satisfied. These fit parameters are not global in nature because tuning of characteristic frequencies perturbs the resonator-resonator coupling and radiative damping rates. Nonetheless, this supports the interpretation of the observed plasmonic spectrum as the result of an EIT-like phenomena rather than a generic strong-coupling effect.

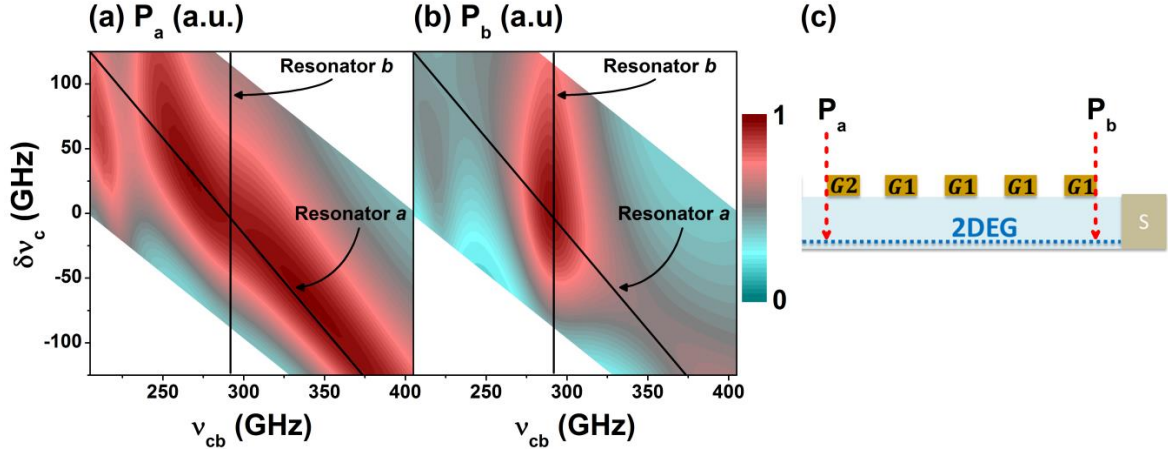

Figure 7. A TL model calculation of the power at the edge of (a) oscillator $a$ and (b) oscillator $b$ with 210.0 GHz excitation is shown for the 2D plasmonic structure diagrammed in Fig. 5. $P_a$ and $P_b$ are plotted as a function of the characteristic frequency $v_{cb}$ of the 'dark' resonator $b$ and its detuning $\delta v_c$ from 'bright' resonator $a$. All parameters used in the model were measured experimentally or are taken from the literature. The color scales in both of the preceding figure parts are adjusted such that deep blue regions are relative minima and dark red regions are relative maxima in power. (c) The locations in the structure where $P_a$ and $P_b$ were calculated are indicated by arrows.

Using the TL model of Fig. 6 (b) to calculate the average power associated with oscillators $a$ and $b$ further validates that this is an EIT-like spectrum. Experimentally, only one component of the complex amplitude of oscillator $a$ is measured using an integrated plasmonic detection mechanism. The amplitude of oscillator $b$ is not directly accessible. However, using the TL model approach, the time-averaged power, $P_a = 1/2\ Re[V_a I_a^*]$ and $P_b = 1/2\ Re[V_b I_b^*]$, is found below the left edge of $G2$ and below far right edge of $G1$, respectively. $P_a$ and $P_b$ are plotted in Figs. 7 (a) and (b), respectively, with the physical locations where these powers were calculated indicated in Fig. 7 (c). As is the case for an EIT spectrum, the in-coupled power to the entire system measured at oscillator $a$ has a broad resonance with a dip near its center. Yet the power spectrum of oscillator $b$ has a single line centered at the dip in oscillator $a$'s spectrum, exactly as illustrated in Figs. 4(c) and (d). This Fano power spectrum arises unambiguously from the interference between the standing plasma waves of these two localized modes.

To contrast these spectra with a 2D plasmonic system featuring more general strong coupling behavior, $P_a$ and $P_b$ of two plasmonic resonators separated by a stripe of ungated 2DEG under 405.0 GHz excitation are plotted in Figs. 8 (a) and (b). The model system is identical to a design studied, both theoretically and experimentally, in current work in the literature (Dyer *et al.*, Phys. Rev. Lett., 109, 126803 (2012))[21]. As illustrated in Fig. 8 (c), two independently tuned 2 μm gate terminals define resonators $a$ and $b$, boxed in blue and red, respectively, while an ungated 2 μm stripe of 2DEG, boxed in green, provides the coupling mechanism. Although in this model the lumped excitation is applied adjacent to oscillator $a$, the power spectrum at oscillator $b$ is virtually identical to that of oscillator $a$ in terms of the location of resonances. This is directly analogous to the scenario plotted in Fig. 4 (b) where two strongly coupled, identically damped oscillators are driven with a large asymmetry. Here, however, there are multiple repelled crossings due to lifting of the degeneracy of multiple higher order resonances that in isolation would have identical characteristic frequencies. The simple proximity of these two oscillators, in contrast to the localized state oscillators in Fig. 7 that are separated by a region that impedes their coupling, requires that when one absorbs resonantly, so does the other. Consequently, there is no mechanism by which oscillator $b$ may resonate and not 'leak' energy through exactly the same loss channels as oscillator $a$. The condition for a Fano or EIT-like spectrum that the two resonators have drastically different damping rates and pathways cannot be satisfied for the system described in Fig. 8. Instead, the power spectra most closely resemble an array of Autler-Townes doublets.

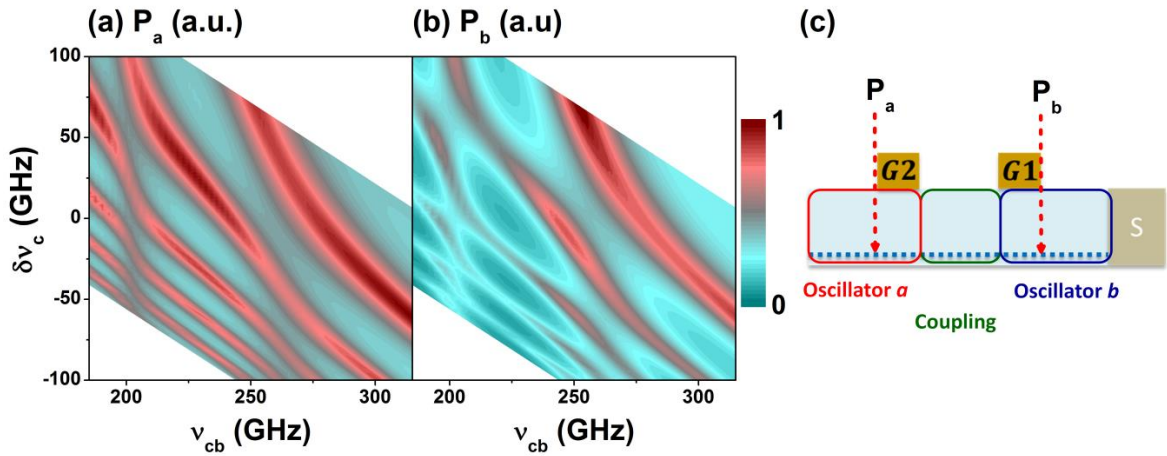

Figure 8. A TL model calculation of the power at the edge of (a) oscillator $a$ and (b) oscillator $b$ with 405.0 GHz excitation is shown for the 2D plasmonic structure diagrammed in Fig. 5. $P_a$ and $P_b$ are plotted as a function of the characteristic frequency $\nu_{cb}$ of resonator $b$ and its detuning $\delta\nu_c$ from resonator $a$. All parameters used in the model were measured experimentally or are taken from the literature. The color scales in both of the preceding figure parts are adjusted such that deep blue regions are relative minima and dark red regions are relative maxima in power. (c) The locations in the structure where $P_a$ and $P_b$ were calculated are indicated by arrows.

## 4. CONCLUSIONS

In this paper, we have elaborated upon a transmission line formalism for describing low dimensional plasmonic systems. Most significantly, we have connected the distributed circuit TL model approach with a general coupled oscillator model. While generic coupled oscillator models are more appropriate for characterizing lumped rather than distributed circuit networks, when localized resonances are present in distributed circuit systems the underlying qualitative features are no different than those of discrete oscillator systems. This paves the way for the rational design and engineering of deeply subwavelength, tunable, coupled 2D plasmonic resonator systems. Because EIT-like resonator-resonator coupling gives rise to slow light, this is an especially promising route to develop frequency agile dispersive photonic elements. The phenomena observed in the device studied in Fig. 5-7 arose without any attempt to optimize device asymmetries for a stronger EIT-like effect. This suggests that there are significant opportunities to enhance the performance of 2D plasmonic coupled oscillator systems based on the principles developed in this paper.

# 5. ACKNOWLEDGEMENTS

The work at Sandia National Laboratories was supported by the DOE Office of Basic Energy Sciences. This work was performed, in part, at the Center for Integrated Nanotechnologies, a U.S. Department of Energy, Office of Basic Energy Sciences user facility. Sandia National Laboratories is a multi-program laboratory managed and operated by Sandia Corporation, a wholly owned subsidiary of Lockheed Martin Corporation, for the U.S. Department of Energy's National Nuclear Security Administration under contract DE-AC04-94AL85000.

# REFERENCES

[1] Smith, D. R., Padilla, W. J., Vier, D. C., Nemat-Nasser, S. C., and Schultz, S., "Composite Medium with Simultaneously Negative Permeability and Permittivity," Phys. Rev. Lett., 84(18), 4184-4187 (2000).
[2] Shelby, R. A., Smith, D. R., and Schultz, S., "Experimental Verification of a Negative Index of Refraction," Science, 292(5514), 77-79 (2001).
[3] Padilla, W. J., Basov, D. N., and Smith, D. R., "Negative refractive index metamaterials," Materials Today, 9(7–8), 28-35 (2006).
[4] Chen, H.-T., Padilla, W. J., Zide, J. M. O., Gossard, A. C., Taylor, A. J., and Averitt, R. D., "Active terahertz metamaterial devices," Nature, 444(7119), 597-600 (2006).
[5] Chen, H.-T., O'Hara, J. F., Azad, A. K., Taylor, A. J., Averitt, R. D., Shrekenhamer, D. B., and Padilla, W. J., "Experimental demonstration of frequency-agile terahertz metamaterials," Nature Photon., 2(5), 295-298 (2008).
[6] Allen, S. J., Tsui, D. C., and Logan, R. A., "Observation of the Two-Dimensional Plasmon in Silicon Inversion Layers," Phys. Rev. Lett., 38(17), 980-983 (1977).
[7] Allen, S. J., Stormer, H. L., and Hwang, J. C. M., "Dimensional resonance of the two-dimensional electron gas in selectively doped GaAs/AlGaAs heterostructures," Phys. Rev. B, 28(8), 4875 (1983).
[8] Tsui, D. C., Gornik, E., and Logan, R. A., "Far infrared emission from plasma oscillations of Si inversion layers," Solid State Comm., 35(11), 875-877 (1980).
[9] Ju, L., Geng, B., Horng, J., Girit, C., Martin, M., Hao, Z., Bechtel, H. A., Liang, X., Zettl, A., Shen, Y. R., and Wang, F., "Graphene plasmonics for tunable terahertz metamaterials," Nature Nano., 6(10), 630-634 (2011).
[10] Fei, Z., Rodin, A. S., Andreev, G. O., Bao, W., McLeod, A. S., Wagner, M., Zhang, L. M., Zhao, Z., Thiemens, M., Dominguez, G., Fogler, M. M., Neto, A. H. C., Lau, C. N., Keilmann, F., and Basov, D. N., "Gate-tuning of graphene plasmons revealed by infrared nano-imaging," Nature, 487(7405), 82-85 (2012).
[11] Chen, J., Badioli, M., Alonso-Gonzalez, P., Thongrattanasiri, S., Huth, F., Osmond, J., Spasenovic, M., Centeno, A., Pesquera, A., Godignon, P., Zurutuza Elorza, A., Camara, N., de Abajo, F. J. G., Hillenbrand, R., and Koppens, F. H. L., "Optical nano-imaging of gate-tunable graphene plasmons," Nature, 487(7405), 77-81 (2012).
[12] Yan, H., Li, X., Chandra, B., Tulevski, G., Wu, Y., Freitag, M., Zhu, W., Avouris, P., and Xia, F., "Tunable infrared plasmonic devices using graphene/insulator stacks," Nature Nano., 7(5), 330-334 (2012).
[13] Yan, H., Li, Z., Li, X., Zhu, W., Avouris, P., and Xia, F., "Infrared Spectroscopy of Tunable Dirac Terahertz Magneto-Plasmons in Graphene," Nano Lett., 12(7), 3766-3771 (2012).
[14] Yan, H., Low, T., Zhu, W., Wu, Y., Freitag, M., Li, X., Guinea, F., Avouris, P., and Xia, F., "Damping pathways of mid-infrared plasmons in graphene nanostructures," Nature Photon., 7, 394-399 (2013).
[15] Di Pietro, P., OrtolaniM, LimajO, Di Gaspare, A., GilibertiV, GiorgianniF, BrahlekM, BansalN, KoiralaN, OhS, CalvaniP, and LupiS, "Observation of Dirac plasmons in a topological insulator," Nature Nano., advance online publication, (2013).
[16] Stern, F., "Polarizability of a Two-Dimensional Electron Gas," Phys. Rev. Lett., 18(14), 546 (1967).
[17] Dyakonov, M. I., and Shur, M. S., "Shallow water analogy for a ballistic field effect transistor: New mechanism of plasma wave generation by dc current," Phys. Rev. Lett., 71(15), 2465 (1993).
[18] Sydoruk, O., Syms, R. R. A., and Solymar, L., "Distributed gain in plasmonic reflectors and its use for terahertz generation," Opt. Express, 20(18), 19618-19627 (2012).
[19] Karabiyik, M., Al-Amin, C., and Pala, N., "Deep Sub-Wavelength Multimode Tunable In-Plane Plasmonic Lenses Operating at Terahertz Frequencies," IEEE Trans. on THz Sci. and Tech., 3(5), 550-557 (2013).
[20] Aizin, G. R., and Dyer, G. C., "Transmission line theory of collective plasma excitations in periodic two-dimensional electron systems: Finite plasmonic crystals and Tamm states," Phys. Rev. B, 86(23), 235316 (2012).

[21] Dyer, G. C., Aizin, G. R., Preu, S., Vinh, N. Q., Allen, S. J., Reno, J. L., and Shaner, E. A., "Inducing an Incipient Terahertz Finite Plasmonic Crystal in Coupled Two Dimensional Plasmonic Cavities," Phys. Rev. Lett., 109(12), 126803 (2012).
[22] Dyer, G. C., Aizin, G. R., Allen, S. J., Grine, A. D., Bethke, D., Reno, J. L., and Shaner, E. A., "Induced transparency by coupling of Tamm and defect states in tunable terahertz plasmonic crystals," Nature Photon., 7(11), 925-930 (2013).
[23] Brews, J. R., "Characteristic Impedance of Microstrip Lines," IEEE Trans. on Microwave Theory and Techniques, 35(1), 30-34 (1987).
[24] Davoyan, A. R., Popov, V. V., and Nikitov, S. A., "Tailoring Terahertz Near-Field Enhancement via Two-Dimensional Plasmons," Phys. Rev. Lett., 108(12), 127401 (2012).
[25] Kachorovskii, V. Y., and Shur, M. S., "Current-induced terahertz oscillations in plasmonic crystal," Appl. Phys. Lett., 100(23), 232108 (2012).
[26] Zhang, S., Genov, D. A., Wang, Y., Liu, M., and Zhang, X., "Plasmon-Induced Transparency in Metamaterials," Physical Review Letters, 101(4), 047401 (2008).
[27] Tassin, P., Zhang, L., Zhao, R., Jain, A., Koschny, T., and Soukoulis, C. M., "Electromagnetically Induced Transparency and Absorption in Metamaterials: The Radiating Two-Oscillator Model and Its Experimental Confirmation," Phys. Rev. Lett., 109(18), 187401 (2012).
[28] Scalari, G., Maissen, C., Turčinková, D., Hagenmüller, D., De Liberato, S., Ciuti, C., Reichl, C., Schuh, D., Wegscheider, W., Beck, M., and Faist, J., "Ultrastrong Coupling of the Cyclotron Transition of a 2D Electron Gas to a THz Metamaterial," Science, 335(6074), 1323-1326 (2012).
[29] Fano, U., "Effects of Configuration Interaction on Intensities and Phase Shifts," Physical Review, 124(6), 1866-1878 (1961).
[30] Luk'yanchuk, B., Zheludev, N. I., Maier, S. A., Halas, N. J., Nordlander, P., Giessen, H., and Chong, C. T., "The Fano resonance in plasmonic nanostructures and metamaterials," Nat. Mater., 9(9), 707-715 (2010).
[31] Garrido Alzar, C. L., Martinez, M. A. G., and Nussenzveig, P., "Classical analog of electromagnetically induced transparency," American Journal of Physics, 70(1), 37-41 (2002).
[32] Tassin, P., Zhang, L., Koschny, T., Economou, E. N., and Soukoulis, C. M., "Low-Loss Metamaterials Based on Classical Electromagnetically Induced Transparency," Phys. Rev. Lett., 102(5), 053901 (2009).
[33] Fleischhauer, M., Imamoglu, A., and Marangos, J. P., "Electromagnetically induced transparency: Optics in coherent media," Reviews of Modern Physics, 77(2), 633-673 (2005).
[34] Anisimov, P. M., Dowling, J. P., and Sanders, B. C., "Objectively Discerning Autler-Townes Splitting from Electromagnetically Induced Transparency," Physical Review Letters, 107(16), 163604 (2011).
[35] Vinogradov, A. P., Dorofeenko, A. V., Merzlikin, A. M., and Lisyansky, A. A., "Surface states in photonic crystals," Physics-Uspekhi, 53(3), 243 (2010).
[36] Geiser, M., Walther, C., Scalari, G., Beck, M., Fischer, M., Nevou, L., and Faist, J., "Strong light-matter coupling at terahertz frequencies at room temperature in electronic LC resonators," Appl. Phys. Lett., 97(19), 191107 (2010).
[37] Muravjov, A. V., Veksler, D. B., Popov, V. V., Polischuk, O. V., Pala, N., Hu, X., Gaska, R., Saxena, H., Peale, R. E., and Shur, M. S., "Temperature dependence of plasmonic terahertz absorption in grating-gate gallium-nitride transistor structures," Appl. Phys. Lett., 96(4), 042105 (2010).